\documentclass{birkjour} 

\usepackage[noadjust]{cite}
\usepackage{xcolor}
\RequirePackage[all]{xy}

\usepackage{amssymb} 
\usepackage{amsfonts} 
\usepackage{amsmath} 
\usepackage{slashed} 

\usepackage{egothic}%
\usepackage{pgothic}%
\usepackage{yfonts}%

 \usepackage{yhmath} 
  
\theoremstyle{definition}

\theoremstyle{remark}

\newcommand{\BibTeX}{B\kern-0.1emi\kern-0.017emb\kern-0.15em\TeX}
\newcommand{\XYpic}{$\mathrm{X\kern-0.3em\raisebox{-0.18em}{Y}}$-$\mathrm{pic}\,$}

\newcommand{\cl}{C \kern -0.1em \ell}  

\newcommand{\ed}{\end{document}}

 \usepackage{yhmath} 
  
\newcommand{\beq}{\begin{equation}}
\newcommand{\eeq}{\end{equation}}
\newcommand{\beqa}{\begin{eqnarray}}
\newcommand{\eeqa}{\end{eqnarray}}
\newcommand{\nn}{\nonumber}
\newcommand{\half}{\frac{1}{2}}

\newcommand{\pbr}[2]{ \{ \hspace*{-2.6pt} [ #1 , #2\hspace*{1.4 pt} ] 
\hspace*{-2.6pt} \} }

\newcommand{\we}{\wedge}
\newcommand{\der}{\partial}

\newcommand{\inn}{\hspace*{2pt}\raisebox{-1pt}{\rule{6pt}{.3pt}\hspace*
{0pt}\rule{.3pt}{8pt}\hspace*{3pt}}}

\newcommand{\ka}{\varkappa}

\newcommand{\Psib}{\overline{\Psi}}
\newcommand{\Phib}{\overline{\Phi}}

\newcommand{\what}[1]{\widehat{#1}}

\newcommand{\bx}{{\mathbf{x}}}

\newcommand{\BPsi}{{\bf \Psi}} 
   
\DeclareMathOperator{\Tr}{Tr} 

\newcommand{\rd}{\mathrm{d}} 
\newcommand{\ri}{\mathrm{i}} 

\newcommand{\omegab}{\bar{\omega}}

\newcommand{\ugamma}{\underline{\gamma}}

\newcommand{\fe}{\mathfrak{e}}

\begin{document}

\title[Precanonical Quantum Teleparallel Gravity]{Towards Precanonical Quantum \\ Teleparallel Gravity}
\author[]{Igor V. Kanatchikov}
\address{%
National Quantum Information Centre KCIK, 81-931 Sopot, Poland \\
IAS-Archimedes Project, C\^{o}te d’Azur, France}
\email{kanattsi@gmail.com}

\begin{abstract}
Quantization of the teleparallel equivalent of general relativity (TEGR) is discussed from the perspective of the space-time symmetric De Donder-Weyl (DW) Hamiltonian formulation with constraints and its quantization called pre-canonical quantization. The representations of operators and the covariant Schr\"odinger equation for TEGR are obtained from the quantization of generalized Dirac brackets calculated according to the analysis of constraints within the polysymplectic formulation of the DW Hamiltonian theory. We argue that the appropriate treatment of the operator ordering and the generalized Hermicity of operators results in an additional $c$-number term in the DW Hamiltonian operator, which is identified with the cosmological constant and estimated to be consistent with its observed value.
\end{abstract}
\label{page:firstblob}
\maketitle
\tableofcontents

\section{Introduction}

The teleparallel description of gravity \cite{per-book,per-rev,maluf-rev} is characterized by the 
vanishing of the curvature tensor and the nonvanishing torsion tensor. 
It corresponds to the  Weitzenb\"ock geometry replacing the Riemann geometry 
of space-time in general relativity. The dynamics of the teleparallel equivalent of general relativity (TEGR) has the same field equations as in general relativity despite the underlying geometry being different.

Teleparallel gravity has been put forward as a potentially favourable framework for the quantization of gravity and reconciliation of gravitation with quantum mechanics \cite{per-q1,per-q2,per-q3}. 
However, the project was never completed satisfactorily despite several attempts in the literature. 
It is not surprising because the potential advantages of teleparallel formulations,  
such as the relation to the gauge theory of the translation group 
and the  ability to define the gravitational energy-momentum tensor 
\cite{per-rev,per-book}
  can not help to resolve the fundamental conceptual and technical problems of quantum gravity 
  which are almost as inevitable in the approaches based on applying the standard methods of quantization to teleparallel gravity
as they are 
 inherent to the approaches starting from general relativity,  viz., 
  the mathematical definition of the (teleparallel analogue of the) Wheeler-De Witt equation,  
the problem of time, the interpretation of (the measurement problem in) quantum cosmology, or the problem of the correct classical limit in (the teleparallel analogue of) 
loop quantum gravity (see e.g. \cite{perim,mielke,warsaw}). 

As a response to these  problems we have put forward an approach to quantization which departs from the various forms of canonical quantization, which are based on the canonical Hamiltonian formalism where a distinguished time variable is required, and instead uses a generalization of the Hamiltonian formalism from mechanics to field theory which treats space and time variables on equal footing. This {\em precanonical quantization\/} \cite{ik2,ik3,ik4,ik5,ik5e} is based on the De Donder-Weyl (DW) Hamiltonian-like formulation in the calculus of variations \cite{kastrup,dw} and a generalization of the Poisson brackets to this formulation found in the earlier papers \cite{mybr1,mybr2,mybr3,ik5}  (see also \cite{helein1,helein2,helein3,mybr-pc,paufler1}). The bracket is defined on differential forms rather than functionals in the canonical formalism, and it leads to the Gerstenhaber algebra structure as a generalization of the Poisson algebra structure in the canonical formalism (see \cite{mybr3,ik5} for more details). 

Quantization of the Heisenberg-like subalgebra of this Gerstenhaber algebra leads to the precanonical 
quantization of fields (see \cite{ik3,ik4,ik5,ik5e} for more details). The result is a hypercomplex generalization of quantum theory where operators and wave functions are Clifford-algebra-valued, 
the Clifford algebra in question is the complexified Clifford algebra of space-time which naturally emerges from the procedure of precanonical quantization of dynamical variables represented by differential forms. 
There is no distinction in this approach between space and time variables and the notion of field configurations and their initial/boundary data, i.e.  sections of the bundle whose base is space-time and whose fibers are spaces where the fields take values, is as superficial as the notion of particle trajectories in quantum mechanics. The wave functions describing quantum fields are (Clifford-algebra-valued) functions on the finite-dimensional bundle of field variables $\phi^a$ over the space-time $x^\mu$:
\beq 
\Psi (\phi^a,x^\mu)= \psi + \psi_\mu\gamma^\mu + \frac{1}{2!}\psi_{\mu\nu}\gamma^{\mu\nu}
+ \frac{1}{3!}\psi_{\mu\nu\rho}\gamma^{\mu\nu\rho} + 
\psi_{0123}\gamma^{0123}
 . 
\eeq 
For example, for the quantum theory of Yang-Mills fields $A_\mu^a$, the precanonical wave function will be $\Psi (A^a_\mu,x^\mu)$ \cite{ik5e,iky1,my-ymmg,iky3} 
and for quantum metric gravity, the precanonical wave function will be $\Psi (g^{\mu\nu},x^\mu)$ \cite{ikm1,ikm2,ikm3,ikm4}. The analogue of the Schr\"odinger equation for the precanonical wave function has the form 
\cite{ikm1,ikm2,ikm3,ikm4,ikv1,ikv2,ikv3,ikv4,ikv5}
\beq  \label{pseq}
i\hbar\varkappa \what{\slashed \nabla} \Psi = \what{H} \Psi ,
\eeq
and $\what{H}$ is the operator of the covariant analogue of the Hamiltonian in the DW Hamiltonian-like formulation of fields,  $\what{\slashed \nabla}$ is the operator of the Dirac operator on the space-time. In the context of quantum gravity, this operator includes tetrads and spin connection coefficients which can be operators themselves.  The parameter $\varkappa$ is an ultraviolet quantity of the dimension of the inverse spatial volume which 
appears on purely dimensional grounds (and  disappears in one-dimensional space-time  that corresponds to quantum mechanics). 

The description of quantum fields in terms of Clifford-algebra-valued
precanonical wave functions on a finite-dimensional space of field variables and space-time variables 
is, obviously, very different from the conventional representations of QFT in terms of functionals of initial field configurations (in the Schr\"odinger representation) or operator-valued distributions 
(in the Heisenberg representation),  or nets of algebras of local observables (in algebraic QFT). 
Nevertheless, we were able to show that the standard functional Schr\"odinger representation of QFT can be derived from the precanonical Schr\"odinger equation (\ref{pseq}) in the limiting case when 
$\varkappa$ goes over to the unregularized 
total volume of the momentum space which 
is $\delta(\mathbf{0})$, the Dirac delta function at equal spatial points \cite{ik-pla,iks1,iks2}. This relation to the standard QFT is proven to hold for scalar fields \cite{iks1,iks2}, Yang-Mills fields \cite{iky3}, and scalar fields in curved space-time \cite{iksc1,iksc2,iksc3}.  In fact, from the probabilistic interpretation of the precanonical wave function - as the probability amplitude of detecting the field value $\phi$ at the space-time point $x$ - and the Schr\"odinger wave functional - as the probability amplitude of observing the field configuration $\phi(\bx)$ on the hypersurface of constant time $t$ - one can anticipate that the  Schr\"odinger wave functional $\BPsi([\phi(\bx)], t)$  is a continuous product (or product integral) over all spatial points $\bx$ of precanonical wave functions restricted to the configuration 
$\phi = \phi(\bx)$, i.e. $\Psi (\phi = \phi(\bx), \bx, t)$. 
At the same time, 
 the latter has to be inserted in the product integral in the form transformed from the representation with the diagonal $\hat{H}$ to the representation with the diagonal energy density operator which has the form 
$\hat{T}{}^0_0 
= \hat{H} - \der_i \phi\, \hat{p}_\phi^i$, where $p_\phi^\mu := \frac{\der L}{\der \der_\mu \phi}$ are the polymomenta and $H := \der_\mu \phi\, p^\mu_\phi - L$  is the classical definition of the DW Hamiltonian function from the Lagrangian $L = L (\phi, \der_\mu \phi, x^\nu)$.   
Let us recall that,  in terms of new Hamiltonian-like variables $p^\mu_\phi$ and $H (\phi, p^\mu_\phi,x^\mu)$,  the Euler-Lagrange equations corresponding to $L$ take the DW Hamiltonian form 
\begin{align}  \label{dwheq}
\der_\mu \phi = \frac{\der H}{\der p_\phi^\mu},  \quad 
\der_\mu p_\phi^\mu =  - \frac{\der H}{\der \phi} , 
\end{align}
provided the Lagrangian function is regular, i.e. 
$\det \big( \frac{\der^2 L}{\der \phi_\mu \der \phi_\nu} \big) \neq 0$. The symbol $\phi$ here may also mean a multicomponent object together with all its internal and space-time indices, such as  
$\phi^a$ or $A_\mu$.
 When the regularity condition is not satisfied, one needs a proper treatment of constraints within the DW Hamiltonian theory. 
 The approach based on the notion of the polysymplectic structure and a Dirac-like generalization of   {Poisson-Gersten\-haber} brackets of differential forms representing both dynamical variables and  constraints has been developed in \cite{mydirac}. Further development and applications of the approach can be found in \cite{ikv1,ikv2,mx1,mx2,mx3,mohj1,mohj2}. Note that a more intrinsically geometric multisymplectic approach to constrained DW Hamiltonian field theories has been elaborated recently in \cite{rr2023,marco22}. Though these authors avoid using the classical Dirac's procedure based on brackets, their analysis of the geometry of constrained DW Hamiltonian systems may help 
 to better understand the geometric foundations of our approach and the reduction 
 of the polymomentum phase space it leads to.  

Note also that the DW Hamiltonian formulation of field dynamics in (\ref{dwheq}) has an associated with it analogue of the Hamilton-Jacobi theory dating back to De Donder and Weyl \cite{dw,kastrup}. In the context of gravity, the latter has been recently studied in \cite{mohj1,riahi}. The derivation of the DW Hamilton-Jacobi equation from the precanonical analogue of the Schr\"odinger equation for scalar fields has been discussed in \cite{ik3,guiding}. A possible extension of this derivation to other fields is yet to be studied. 
 
 Here, we will explore the precanonical quantization of the teleparallel equivalent of tetrad general relativity. To this end, in Section 2, we first present a proper Palatini formulation of TEGR and its DW Hamiltonian formulation, which will turn out to be constrained. The analysis of constraints and the calculation of the generalized Dirac brackets allow us to identify the variables of the reduced polymomentum phase space. Then, in Section 2.3, the quantities of the theory, such as the DW Hamiltonian function, are expressed as functions of the reduced polymomentum phase space. 
 In Section 3, we quantize the resulting DW Hamiltonian system on the reduced polymomentum phase space according to the  modified Dirac quantization rule and find the representations of operators up to an operator ordering ambiguity. In Section 3.2, we formulate the covariant precanonical Schr\"odinger equation for TEGR, and then, in Section 3.4, we discuss the scalar product of Clifford-valued precanonical wave functions. In the final Section 3.5, we discuss the problem of operator ordering consistent with the weight factor in the integration measure in the scalar product and, as a consequence, estimate the constant which results from the proper operator ordering in the DW Hamiltonian function and its relation with the observable value of the cosmological constant.

\section{Palatini tetrad TEGR} 

The teleparallel description of general relativity \cite{per-book,per-rev,maluf-rev} is characterized by the vanishing curvature tensor and the nonvanishing torsion tensor. It corresponds to the  Weitzenb\"ock geometry replacing the Riemann geometry of space-time in general relativity. The field equations of the teleparallel equivalent of general relativity (TEGR) are the same field equations of general relativity despite the underlying geometry being different. 

In our previous papers \cite{ikv1,ikv2,ikv3,ikv4,ikv5}, we have presented a procedure of precanonical quantization of vielbein general relativity in the Palatini formulation. Here we will show how this approach can be extended to the teleparallel equivalent of general relativity (TEGR) in the Palatini tetrad formulation. Though the related formulations of TEGR have been considered in many papers (see, e.g., \cite{maluf2,koivisto}), this is the Palatini formulation found by Maluf \cite{maluf} we found to be the most appropriate one. 
 
In this formulation,  the  Lagrangian density 
\beq  \label{lagrtp}
\mathfrak{L} = \frac{1}{16\pi G} \fe \Phi (f)^{abc} \left(f_{abc}-2T_{abc}\right)
\eeq
uses tetrads $e^\mu_a$ and the auxiliary variables $f_{abc}= - f_{acb}$ 
 with the last two antisymmetric indices 
as independent field variables. The Gothic letters are reserved for densities, e.g. 
 $\fe := \det (e^a_\mu)$,  
\begin{align}
\Phi (f)_{abc}&:=\frac14 (f_{abc}+f_{bac}-f_{cab}) + \eta_{a[c} f_{b]} ,  \nn \\ 
f_b&:= \eta^{ca}f_{cab}, \nn  
\end{align}
for linear functions of field variables $f_{abc}$,  and 
\begin{align}\label{tors}
\begin{split}
T^{c}{}_{ \mu\nu} &:= 
 2 \der_{[\mu} e^c_{\nu ]} , 
\\
T^c{}_{ab} &:= e^\mu_a e^\nu_b T^c{}_{\mu\nu} , 
\quad 
T_b := T^a{}_{ab}, 
\end{split}
\end{align}
for linear functions of the first jets of tetrad components: $\der_\mu e^c_\nu$. 
The restriction of the Lagrangian density to a field configuration 
$e^c_\nu =  e^c_\nu (x)$, 
$\der_\mu e^c_\nu = \der_\mu e^c_\nu (x) $, $f_{abc} = f_{abc} (x) $ 
and integration over $x^\mu$, $\mu = 0,1,2,3,$ yields the action functional. 
Its variation with respect to the auxiliary fields $f_{abc} (x)$ and 
tetrad fields $e^c_\nu (x)$ leads to the following Euler-Lagrange field equations 
\begin{align} \label{deltaf}
\delta f:& \; \Phi (f)^{abc} = \Phi (T)^{abc} , 
\\
\delta e:& \; e_{a\sigma}e_{b\mu}\der_\tau \left(\fe \Phi^{b\sigma\tau}(T)\right) 
- \fe  \big( \Phi^{b\sigma}{}_a(T)  T_{b\sigma \mu} 
- \frac14 e_{a\mu} T_{cbd} \Phi^{cbd}(T)\big) =0 . 
\label{deltae}
\end{align}
The first equation is equivalent to \cite{maluf94}
\beq  \label{feqt}
f_{abc} = T_{abc}  , 
\eeq
i.e. the auxiliary fields $f_{abc}$ are identified with the torsion tensor (\ref{tors}). 
The second one uses (\ref{feqt}) and reproduces the tetrad form of the vacuum Einstein equations \cite{maluf94}
\beq
R^a_{\mu} - \frac12 e^a_{\mu} R = 0 .
\eeq

\subsection{The DW Hamiltonian analysis}


From (\ref{lagrtp}) we obtain the polymomenta densities:  
\begin{align}
\mathfrak{p}{}^\mu_{f_{abc}}&:= \frac{\der \mathfrak{L}}{\der \der_\mu f_{abc}}=0, 
 \\
\mathfrak{p}{}^\mu_{e^a_\nu} &:= \frac{\der \mathfrak{L}}{\der \der_\mu e^a_\nu}
  =  - \frac{1}{4\pi G}\fe \Phi(f)_a{}^{bc} e^{[\mu}_b e^{\nu]}_c , 
\end{align}
 the resulting primary constraints:
\begin{align}
\mathfrak{C}{}^\mu_{f_{abc}} & :=\mathfrak{p}{}^\mu_{f_{abc}} \approx 0, \label{cf}\\
 \mathfrak{C}{}^\mu_{e^a_\nu}\; & :=  \mathfrak{p}{}^\mu_{e^a_\nu} +  \frac{1}{4\pi G}\fe \Phi(f)_a{}^{bc} e^{[\mu}_b e^{\nu]}_c \approx 0 , 
\label{ce}
\end{align}
and the primary DW Hamiltonian density: 
\beq \label{dwhprim}
{\fe H} := \mathfrak{p}^\mu_f \der_\mu f + \mathfrak{p}^\mu_e \der_\mu e - \mathfrak{L} 
\approx  - \frac{1}{16\pi G} \fe \Phi (f)^{abc} f_{abc} 
\approx \frac14 \mathfrak{p}^\mu_{e^a_\nu} e_\mu^b e_\nu^c f^a{}_{bc}  .
\eeq
Note that here and in what follows we adopt a convention that the symbol of an object serves as a collection of all its indices, e.g., 
$\mathfrak{p}^\mu_f \der_\mu f := 
\mathfrak{p}^\mu_{f_{abc}} \der_\mu f_{abc}$ and $\mathfrak{p}^\mu_e \der_\mu e := \mathfrak{p}^\mu_{e^a_\nu} \der_\mu {e^a_\nu}$. 

Now, the unconstrained (extended) polymomentum phase space of variables \
\mbox{$(e, f, \mathfrak{p}^\mu_e, \mathfrak{p}^\mu_f, x^\mu)$} is naturally~equip\-ped with the (pre)polysymplectic structure represented by 
(a representative of the equivalence class of forms modulo semi-basic $4-$forms \cite{ik5}) 
\beq \label{preps}
\Omega = d\mathfrak{p}{}^\mu_e \we d e \we \varpi_\mu 
+ d \mathfrak{p}{}^\mu_f \we d f \we \varpi_\mu , 
\eeq
where $\varpi_\mu := \der_\mu \inn (dx^0\we dx^1\we dx^2 \we dx^3)$. 
As the object $\Omega$ maps (a certain class of) semi-basic 
$p-$forms $F$ for $0 \leqslant p \leqslant 3$ to (an equivalence class of) $(4-p)-$multivector fields $X_F$  (modulo the multivectors $X_0$ from the kernel: \mbox{$X_0 \inn\ \Omega =0$}), i.e.  
\beq  \label{map}
X_F \inn\ \Omega = d F , 
\eeq
it defines a bracket operation \cite{mybr1,mybr2,mybr3,ik5} 
\beq \label{gbr}
\pbr{F}{G} := (-1)^{p} X_F\inn\ dG  . 
\eeq
This bracket leads to the Poisson-Gerstenhaber structure on the space of semi-basic 
forms  equipped with the co-exterior product operation $\bullet$: 
\beq \label{}
F\bullet G := *^{-1}(*F\we*G) , 
\eeq
which preserves the space of forms for which the map (\ref{map}) exists (called Hamiltonian forms in my first paper on the subject \cite{mybr1}; perhaps, a more appropriate term would be ``admissible forms", cf. \cite{courant}). Note that a more general class of ``Poisson forms" and their brackets are defined in \cite{paufler1}. However, there is no product operation on the space of those forms, so the corresponding bracket operation is graded Lie, i.e. not a   generalization of the Poisson bracket. 

Using the definition above we can calculate the brackets of $3-$forms of  constraints  which were first introduced in \cite{mydirac} 
\beq 
\mathfrak{C}{}_e:=  \mathfrak{C}{}^\mu_{e}\varpi_\mu,  \quad 
\mathfrak{C}{}^\mu _f:= \mathfrak{C}{}^\mu_{f}\varpi_\mu .
\eeq
We obtain  
\beqa
&\pbr{\mathfrak{C}{}_f}{\mathfrak{C}{}_{f'}} &\!\!\! =: \mathfrak{C}{}_{f f'} = 0 , \\
&\pbr{\mathfrak{C}{}_{e^a_\mu}}{\mathfrak{C}{}_{e^d_\nu}} &\! \!\! = \frac{1}{4\pi G} \der_{e^a_\mu}\left( \fe \Phi(f)_d{}^{bc} e^{[\alpha}_b e^{\nu]}_c \right)
\varpi_\alpha =: \mathfrak{C}{}_{e^a_\mu e^d_{\nu}}
 \quad \mathrm{or}\quad \mathfrak{C}{}_{e e'}  , 
\label{cece} \\
&\pbr{\mathfrak{C}{}_{e^d_\nu}}{\mathfrak{C}{}_{f_{abc}}} &\!\!\! = -\frac{1}{4\pi G}\fe e^{[\mu}_g e^{\nu]}_h 
\frac{\der\Phi(f)_d{}^{gh}}{\der {f_{abc}}} \varpi_\mu
=: \mathfrak{C}{}_{e^d_\nu f_{abc}}
\quad \mathrm{or}\quad \mathfrak{C}{}_{e f}  . 
\eeqa
Note that $ \mathfrak{C}{}_{e e'} $ depends on $e$ and $f$ whereas $ \mathfrak{C}{}_{e f} $ depends only on $e$.

The block structure of the matrix of brackets of constraints $\mathfrak{C}=\mathfrak{C}_{UV}^\mu\varpi_\mu$, 
whose matrix elements are $3-$forms,  takes the form 
\beq  \mathfrak{C}_{UV} = 
\begin{Vmatrix}   
    \mathfrak{C}_{e e'}       & \mathfrak{C}_{e f} \\
    \mathfrak{C}_{f e} = -  \mathfrak{C}_{e f}^T     & \mathfrak{C}_{f f'} = 0 
\end{Vmatrix} .  
\eeq
The indices $U,V$  enumerate the constraints and run over all the values of $e$ and $f$, i.e. they are $(16+24)=40-$dimensional. 

\subsection{Generalized Dirac brackets and the polysymplectic reduction}

The generalized Dirac brackets of forms $A$ and $B$ on the subalgebra of forms of degree 
$3$ and $0$ has the form \cite{mydirac,ikv1 } 
\beq \label{pdir}
\pbr{A}{B}{}^D = \pbr{A}{B}{} - \pbr{A}{\mathfrak{C}_U}{}\bullet ( \mathfrak{C}_{UV}^{\sim 1} \we \pbr{\mathfrak{C}_V}{B}{}), 
\eeq
where $\mathfrak{C}^{\sim 1} =\mathfrak{C}^{\sim 1}_\mu dx^\mu $ is the generalized pseudoinverse matrix 
whose matrix elements are one-forms  such that (cf. \cite{penrose})
\beq
\mathfrak{C}^{}\bullet \mathfrak{C}^{\sim 1} \!\we \mathfrak{C}^{} = \mathfrak{C}^{}. 
\eeq 
The appearance of the pseudoinverse in our definition of generalized Dirac brackets instead of the inverse matrix in the original definition of Dirac brackets is related to the nonvanishing multivector kernel of the polysymplectic forms while the vector kernel of symplectic forms is vanishing.

The block structure of this generalized pseudoinverse has the form 
\beq\label{pseinv}
\begin{Vmatrix}  
    \mathfrak{C}^{\sim 1}_{e e'}=0    & \mathfrak{C}^{\sim 1}_{e f}   \\
    \mathfrak{C}_{f e}^{\sim 1}= - \mathfrak{C}^{\sim 1}_{e f}{}^T     & \mathfrak{C}^{\sim 1}_{f f'} \! 
\end{Vmatrix}, 
\eeq
with 
\beq  
\mathfrak{C}^{\sim 1}_{\mu e f} \mathfrak{C}^{\mu}_{f e'} = \delta_{ee'}, 
\quad 
\mathfrak{C}^{\sim 1}_{\mu fe } \mathfrak{C}^{\mu}_{ef'} = \delta_{ff'},
\quad 
\mathfrak{C}^{\sim 1}_{\mu f e'} \mathfrak{C}^{\mu}_{e'e}  + \mathfrak{C}^{\sim 1}_{\mu f f'} \mathfrak{C}^{\mu}_{f'e}=0  . 
 \label{cinver}
\eeq
Note that $\mathfrak{C}_{UV}$ are densities of weight $+1$ and  $\mathfrak{C}^{\sim 1}_{UV}$ are \mbox{densities of weight $-1$.} 

For the most relevant brackets we obtain 
\begin{align}
\pbr{\mathfrak{p}{}_f}{f'}{}^D &= \pbr{\mathfrak{p}{}_f}{f'}{} - \pbr{\mathfrak{p}{}_f}{ \mathfrak{C}_e}\bullet (  \mathfrak{C}^{\sim 1}_{e f''} 
\we \pbr{\mathfrak{C}_{{f''}}}{f'})
\nn \\
& \quad = 
\delta_{ff'} - \mathfrak{C}_{fe}\bullet \mathfrak{C}^{\sim 1}_{e f''} \delta_{f'' f'} 
= \delta_{ff'} - \mathfrak{C}^\mu_{fe} \mathfrak{C}^{\sim 1}_{\mu e f'}
= 0,  \!\!\mbox{\!\!} \label{pff}
\\ 
\pbr{\mathfrak{p}{}_f}{e}{}^D & 
= \pbr{\mathfrak{p}{}_f}{\mathfrak{p}{}_{f'}}{}^D 
= \pbr{\mathfrak{p}{}_f}{\mathfrak{p}{}_{e}}{}^D 
= \pbr{\mathfrak{p}{}_e}{\mathfrak{p}{}_{e'}}{}^D =0 , 
\label{pepe}
\\
\pbr{\mathfrak{p}{}_e}{e'}{}^D &= 
 \pbr{\mathfrak{p}{}_e}{e'}{}=\delta_{ee'}  , 
\label{pee}
\\
\pbr{\mathfrak{p}{}^\mu_e}{e'\varpi_\nu}{}^D 
& =  \pbr{\mathfrak{p}{}^\mu_e}{e'\varpi_\nu}{}=\delta_{ee'}\delta^\mu_\nu  , 
\label{pmuee}
\\
\pbr{\mathfrak{p}{}_e}{f}{}^D &=
\pbr{\mathfrak{p}{}_e}{f}{} -  \pbr{\mathfrak{p}{}_e}{\mathfrak{C}_{e'}} \bullet ( \mathfrak{C}^{\sim 1}_{e' f'} \we \pbr{\mathfrak{C}_{f'}}{f} ) 
\nn \\ & \quad 
= - \frac{1}{4\pi G} \der_e \left(\fe{} \Phi(f)_b{}^{cd} e{}_c^\alpha e_d^\nu \right)  \mathfrak{C}^{\sim 1}_{\alpha e^b_\nu f} = 
- \mathfrak{C}^\alpha_{e e'} \mathfrak{C}^{\sim 1}_{\alpha e' f}  , 
\label{pef}\\
\pbr{e \varpi_\mu}{f_{}}^D &
= \mathfrak{C}^{\sim 1}_{\mu ef} , 
\label{ef}
\\
\pbr{e \varpi_\mu}{e'_{}}^D &
=  \mathfrak{C}^{\sim 1}_{\mu ee'} = 0 , 
\\
 \pbr{f \varpi_\mu}{f'_{}}^D &
 = \mathfrak{C}^{\sim 1}_{\mu ff'} . 
 \label{ff}
\end{align}

From these brackets, it follows that the constrained (reduced) polymomentum phase space is spanned by the configuration space 
coordinates $e^a_\mu$ and their conjugate polymomenta $\mathfrak{p}{}^\mu_{e^\nu_a}$ alone, while the  
variables $f_{abc}$, whose different components do not commute according to ({\ref{ff}}),
become functions of the fundamental variables of the reduced polymomentum phase space 
equipped with the reduced polysymplectic structure given by 
\beq \label{omegar}
\Omega_R = d\mathfrak{p}{}^\mu_e \we d e \we \varpi_\mu . 
\eeq  
 {
In should be noted, in order to avoid confusion, that different notions of the polysymplectic structure and polysymplectic reduction exist in the literature \cite{mcclain,rr-psred,blacker}, all of them are inspired by G\"unther's work in the 1980-es \cite{guenther}, including our work starting from \cite{mybr1}.  A comparison of different related geometric structures in classical field theory 
can be found in \cite{deleon-book}. Nevertheless, it is our understanding of the polysymplectic structure as a certain equivalence of class of forms \cite{mybr2,ik5} that allows to define 
a proper analogue of the Poisson brackets in the DW Hamiltonian theory and its Dirac generalization in constrained  DW Hamiltonian theory, 
which allow to quantize field theories. Other proposed generalizations, which are usually similar to the one discussed in \cite{my-dkp},  usually do not have natural 
Leibniz and Jacobi properties which are essential for the possibility of Dirac's quantization 
by replacing the classical bracket with quantum commutator. 
}

\subsection{$f_{abc} (e,\mathfrak{p}{}_e)$   and $H(e,\mathfrak{p}{}_e)$} 

We have to express $f_{abc}$ as a function of $e$ and $\mathfrak{p}{}_e$. From (\ref{ce}) 
it follows that it is possible if the map $f \rightarrow \Phi$ is invertible. 
From eq. (2), we obtain 
\begin{align}
\begin{split}
f_{abc} + \eta_{ac} f_b -\eta_{ab} f_c  &= 2 \Phi_{bac} - 2 \Phi_{cab} , \nn \\
f_c &= 
-  \Phi_c , \nn 
\end{split}
\end{align}
so that 
\beq
\frac12 f_{abc} =  \Phi_{bac} -  \Phi_{cab} 
- \frac12 (\eta_{ab}\Phi_c - \eta_{ac}\Phi_b) . 
\eeq
Using (\ref{ce})  
and the notation ${p}{}_e^\mu := \fe{}^{-1} \mathfrak{p}^\mu_e$ we obtain 
\begin{align}
\begin{split}
 \Phi_{a}{}^{bc} &\approx {-4\pi G}\, e_{ [\mu }^b e^c_{ \nu  ]} {p}{}^\mu_{e^a_\nu} ,  \\
  \Phi^c &= \delta^a_b \Phi_{a}{}^{bc} \approx {-4\pi G}\, e^a_{[\mu} e^c_{\nu ]} p^\mu_{e^a_\nu} . 
\end{split}
\end{align}
We, therefore, obtain 
\begin{align}  \label{fabc}
\begin{split}
 f_{abc} &\approx {8\pi G} \Big(   e_{a [\mu } e_{b| \nu  ]} p^\mu_{e^c_\nu} -   e_{a [\mu } e_{c| \nu  ]} p^\mu_{e^b_\nu}
  \\  & \hspace{50pt} + 
 \frac12 \left( \eta_{a b} e^d_{[\mu} e_{c|\nu ]} p^\mu_{e^d_{\nu}} 
  - \eta_{a c} e^d_{[\mu} e_{b |\nu ]} p^\mu_{e^d_{\nu}} \right) \Big)  .
  \end{split}
\end{align}
The latter can be rewritten in a shorter form 
\beq \label{fabc2}
f_{abc} \approx 
 {16\pi G} \Big(   
 e_{a \mu } e_{[b \nu } {p}{}^{[\mu}_{e^{c]}_{\nu]}}  
+ 
\frac12 \eta_{a [b} e^d_{\mu} e_{c] \nu } p^{[\mu}_{e^d_{\nu]}} 
   \Big)  . 
\eeq

The DW Hamiltonian density $\fe H$  (\ref{dwhprim}) can now be expressed as a function of the variables of the reduced polymomentum phase space: 
\begin{align}
\label{hdwr}
 H  &\approx \frac14 p^\mu_{e^a_\nu} e_\mu^b e_\nu^c f^a{}_{bc}  \nn \\
& \approx 
 {2\pi G}  p^\alpha_{e^a_\beta} e_\alpha^b e_\beta^c \left( -   e_{ [\mu }^a e_{c|\nu]} 
 p^\mu_{e^b_\nu}
+  e_{ [\mu }^a e_{b |\nu  ]} p^\mu_{e^c_\nu}  
+  
   \eta^a_{b} e^d_{[\mu} e_{c |\nu ]} p^\mu_{e^d_\nu}\right)
\nn \\
& \approx {4\pi G}  p^\alpha_{e^a_\beta} e_{[\alpha}^b e_{\beta]}^c \left(
  e_{ [\mu }^a e_{b |\nu  ]} p^\mu_{e^c_\nu}  
+  
\frac{1}{2}  \eta^a_{b} e^d_{[\mu} e_{c |\nu ]} p^\mu_{e^d_\nu} 
\right) . 
\end{align}

Using the notation $p^a_{e^b_\nu} := e_\mu^a p^\mu_{e^b_\nu},  \;
p_{\mu e^b_\nu} = g_{\mu\nu}p^\nu_{e^b_\nu}  $  we can write 
\beq
\begin{aligned}
H  &\approx {\pi G}\Big( 2 e^c_\beta p_{\nu e^a_\beta} p^a_{e^c_\nu } 
-g_{\beta\nu} p^c_{e^a_\beta}p^a_{e^c_\nu}
-g_{\alpha\mu}e^c_\beta e^a_\nu p^\alpha_{e^a_\beta}p^\mu_{e^c_\nu} \\ 
& +  
\frac{1}{2}
\big( 
g_{\beta\nu}p^a_{e^a_\beta}p^b_{e^b_\nu} - 2 e^a_\beta p_{\nu e^a_\beta} p^b_{e^b_\nu}
  + g_{\alpha\mu}e^b_\nu e^a_\beta p^\alpha_{e^a_\beta}p^\mu_{e^b_\nu} 
\big) \Big)  .
\end{aligned} 
\eeq

Let us note here that a mechanical analogue of this Hamiltonian function would be a Hamiltonian of a particle 
with a coordinate-dependent tensor mass: 
\beq
h = \frac12 m^{-1}_{ij}(q) p^i p^j.  
\eeq
A one-dimensional version of such Hamiltonian has been discussed in the context of quantum gravity  by Aslaksen and Klauder already in 1970 \cite{ak}.


\section{Quantization}

Quantization of fundamental brackets (\ref{pff})-(\ref{ff}) should be performed according to the modified Dirac quantization rule
\beq
[\hat{A},\hat{B}] = -i\hbar \what{\mathfrak{e} \pbr{A}{B}^{ D}} .
\eeq 
The modification guarantees that densities are represented by density-valued operators. 
From (\ref{pee})  we can easily obtain that 
\beq  \label{pmuomop}
\what{\mathfrak{p}{}_e^\mu \varpi_\mu} = -i \hbar\fe \der_e .
\eeq 

Then from (\ref{pmuee})  it follows 
\beq \label{pmueop}
\what{\mathfrak{p}}{}_e^\mu  = -i\hbar\ka \fe \gamma^\mu \der_e   
\eeq
and 
\beq \label{omegaop}
\what{\varpi}_\mu = \frac{1}{\varkappa }\gamma_\mu , 
\eeq
provided the composition of Clifford-algebra-valued operators implies the symmetrized Clifford product (see \cite{ik3,ik5e} for detailed explanation). 
The ultra-violet parameter $\varkappa$ is introduced in (\ref{omegaop}) and appears on dimensional grounds, which is a standard element of precanonical quantization. Obviously, $\frac{1}{\varkappa }$ has the dimension and the meaning of a small or minimal spatial volume. 
The curved Dirac matrices 
\beq 
\gamma^\mu := e^\mu_a \ugamma^a  
\eeq 
are defined in terms of the flat Dirac matrices 
$\ugamma^a$ such that 
$$\ugamma^a\ugamma^b+  \ugamma^b\ugamma^a = 2 \eta^{ab}.$$  
Note that the operators in (\ref{pmuomop}), (\ref{pmueop}) are defined only up to an ordering of multiplicative and differential operators, and their precise definition depends on the scalar product to be found (see Section  3.4). 

\subsection{$\what{f}_{abc}$ and $\what{H}$ }

Using the representation of the operators of polymomenta $\hat{p}{}^\mu_{e^a_\nu} $ 
in the classical expression (\ref{fabc}) 
we obtain up to an ordering 
\begin{align}\label{fabcop}
\begin{split}
\hat{f}_{abc} & = - {8 \pi i G} \hbar\ka \Big( \big( \ugamma{}_a e_{[b|\nu} 
- e_{a\nu}\ugamma{}_{[b}\big)  \der_{e^{c]}_\nu} 
\\ &  \hspace{20pt}+ 
 \frac12 
 \big(\ugamma^d \eta_{a[b}e_{c]\nu} 
- \eta_{a[b}\ugamma{}_{c]} e^d_\nu   \big) \der_{e^d_\nu} \Big) . 
\end{split}
\end{align}

From (\ref{hdwr}), the operator $\what{H}$ has the form 
\begin{align}
 \what{H} 
& = 
 {4\pi G} \fe^{-2}  e_{[\alpha}^b e_{\beta]}^c \hat{p}{}^\alpha_{e^a_\beta} \left( 
 e_{ [\mu }^a e_{b |\nu  ]} \hat{p}{}^\mu_{e^c_\nu}  
+ 
    \frac12 
\eta^a_{b} e^d_{[\mu} e_{c |\nu ]} \hat{p}{}^\mu_{e^d_\nu}
\right) . 
\end{align}
Up to an ordering of operators, we obtain 
\begin{align} \label{hop}
 \what{H} = -\pi \hbar^2\ka^2 G  \Big( e^c_\beta e^a_\nu - 
 \frac32 
 g_{\beta\nu} \eta^{ac}
- 
   \frac12 
   e^a_\beta e^c_\nu \Big)
 \der_{e^a_\beta}\der_{e^c_\nu}  .
\end{align}
The derivation uses the identity 
\beq
\gamma^\mu\ugamma{}_a +\ugamma{}_a \gamma^\mu = 2 e^\mu_a   
\eeq
as well as the observation that the operator of $g_{\mu\nu} p^\mu_e p^\nu_{e'}$  takes the form 
$-\hbar^2\ka^2\der_e\der_{e'}$  similarly to the operator of 
$g_{\mu \nu} p^\mu_\phi p^\nu_{\phi}$  in scalar field theory, 
which equals $-\hbar^2\ka^2\der_{\phi\phi} $
\cite{ik3,ik4,ik5e}.


\subsection{The precanonical Schr\"odinger equation for TEGR} 

Precanonical Schr\"odinger equation in curved space-time has the form (cf. \cite{iksc1,iksc2,iksc3,ikm1,ikm2,ikm3,ikm4,ikv1,ikv2,ikv3,ikv4,ikv5}) 
\beq 
\ri \hbar\ka \what{\slashed\nabla}  \Psi \!=\! \what{H} \hspace*{-0.0em} \Psi . 
\eeq 
In the case of quantum TEGR 
$\what{\slashed\nabla}$ denotes   the quantized Dirac operator in curved space-time in which 
 the spin connection is an operator acting via the commutator Clifford product on the 
 Clifford-algebra-valued wave function $\Psi(e,x)$, 
 \beq
 \what{\slashed\nabla} := \ugamma^a e^\mu_a \left(\der_\mu+   
  \what{\omega}_{\mu}\mbox{$\stackrel{\leftrightarrow}{\vee}$}
 \right) , 
 \eeq 
where  $ \what{\omega}_{\mu} =  \frac14 \what{\omega}_{\mu bc}{\ugamma}^{bc}$ 
is the operator of spin connection 
and $\stackrel{{\leftrightarrow}}{\vee}$   denotes the commutator 
(antisymmetric) Clifford product 
\beq
{\ugamma}^{bc} \stackrel{\leftrightarrow}{\vee} \Psi := \frac12\left( {\ugamma}^{bc} \vee \Psi - \Psi \vee {\ugamma}^{bc} \right) = \frac12 \left[\ \ugamma^{bc}, \Psi\ \right] .
\eeq
Here, $\vee$ denotes the Clifford algebra product which is just the matrix product 
denoted by the juxtaposition of matrices as in the second equality. 

  The conjugate wave function  
  \beq
  \Psib := \ugamma{}^0\Psi^\dagger\ugamma{}^0    
  \eeq 
satisfies the equation  
\beq \label{pse-conj}
\ri\hbar\ka \Psib 
{\left( \stackrel{\leftarrow}{\der_\mu}  
+ \mbox{ $\stackrel{\leftrightarrow}{\vee}$} \what{\omegab}{}_\mu\right) } \ugamma^a e^\mu_a
= - \hat{H} \Psib , 
\eeq
where 
$\what{\omegab}_\mu \!:=\!\ugamma{}^0\what{\omega}_\mu^\dagger\ugamma{}^0$,  
if a generalized Hermicity of $\hat{H}$ is assumed: 
 $\hat{H} = {\overline{\!\!\hat{H}\!}}$, where \;
 ${\overline{\!\!\hat{H}\!}} := \ugamma{}^0\hat{H}{}^\dagger\ugamma{}^0$.

 \subsection{The spin connection operator $\what{\omega}_{\mu ab}$}
 
 The operator of spin connection coefficients is a linear combination of operators of $f_{abc}$ found in (\ref{fabcop}). 
 To construct it, let us recall the classical (globally Lorentz covariant) expression
  which relates the teleparallel spin connection to the contorsion \cite{pereira01,maluf03} 
 \beq
 \omega_{\mu ab} = - K_{\mu ab} = \half e^c_\mu\left(T_{a bc} + T_{ac b} - T_{c ab}\right) .
 \eeq
 This is an on-shell expression of the spin connection in terms of the derivatives of tetrads. 
 For quantum representation, we need an off-shell expression in terms of the canonical variables. 
 The field equation (\ref{deltaf}) indicates that on-shell $f_{abc} = T_{abc}$, 
 Therefore, the operator of spin connection coefficients is expressed in terms of operators 
 $\hat{f}_{abc}$: 
 \beq \label{spomegaop}
 \hat{\omega}_{\mu ab} = \half e^c_\mu\left(   \hat{f}_{a bc} + \hat{f}_{ac b} - \hat{f}_{c ab}\right) .
 \eeq
 Obviously, this expression of $\hat{\omega}_{\mu ab}$ also depends on the ordering of $e$ and $\der_e$. 
 
 \subsection{The scalar product} 
 
 This scalar product of Clifford-algebra-valued precanonical wave functions is given by 
\beq \label{scpr}
\left\langle \Phi | \Psi \right\rangle 
:=  \Tr \int\! {[\rd e]}\ \Phib \,  \Psi , \quad 
\eeq 
where 
 ${[\rd e]}$ is a Misner-like 
 diffeomorphism invariant measure 
 on fibres of the configuration bundle of tetrads over space-time: 
\beq \label{measure}
{[\rd e]} 
 :=
 {\mathfrak e}{}^{- 4}\prod_{\mu, a} \rd e_\mu^a .
\eeq
 Then the expectation values of operators are calculated according to 
\beq
 \langle \hat{O}\rangle(x) = 
 \Tr \int\! {[\rd e]}\  \Psib (e,x) 
 \hat{O}
 \Psi (e,x) . 
\eeq
Note that the norm  $\left\langle \Psi | \Psi \right\rangle $ 
corresponding to (\ref{scpr}) is not positive definite. It means that a subspace of physical 
precanonical wave functions has to be singled out by additional conditions. 
For example, the projection $\Psi_+:=\frac14(1-\ugamma^0)\Psi(1-\ugamma^0) $ 
with $(\ugamma^0)^2=1$ satisfies $\Tr (\overline{\Psi}_+ \Psi_+) = \Tr (\Psi_+^\dagger \Psi_+) >0$. 

 \subsection{Operator ordering and the cosmological constant}
 
Because of the weight factor ${\mathfrak e}{}^{- 4}$ in the integration measure (\ref{measure}), the requirement of 
the (generalized) Hermicity of operators will produce additional terms in the formal expressions which we have written up to an ordering in (\ref{pmuomop}), (\ref{pmueop}), (\ref{hop}), (\ref{fabcop}) and (\ref{spomegaop}). 
 For example, the Hermitian version of the ordering-dependent operator (\ref{pmuomop}) has the form 
\beq
\hat{\mathfrak{p}}{}_{e^a_\nu} := 
\what{\mathfrak{p}{}^\mu_{e^a_\nu} \varpi_\mu} = -i\hbar\ka \mathfrak{e} \der_{e^a_\nu} + 
  \frac{3}{2}i\hbar\ka \mathfrak{e} e_a^\nu .   
\eeq
One can check that 
\beq
\Tr\!\int {[\rd e]}\  \Psib (e,x) 
 (\hat{\mathfrak{p}}{}_{e^a_\nu} \Psi (e,x)) = 
\Tr\! \int {[\rd e]}\  (\overline{\hat{\mathfrak{p}}{}_{e^a_\nu}\Psi (e,x)}) \Psi (e,x) . 
\eeq
Similarly, the Hermitian version of the ordering-dependent operator $\hat{H}$ in (\ref{hop}) acquires additional terms which include a constant that emerges from the re-ordering of the term $\frac12 (e \der_e)^2$ in (\ref{hop}). This constant is of the order of $\lambda \sim \frac12 4^4 \pi\hbar^2\varkappa^2 G$, and it appears as the quantum-gravitational contribution to the cosmological constant ($\Lambda/8\pi G$  to be precise). The emerging cosmological constant depends on the scale of the parameter $\varkappa$. The latter appears in precanonical quantization as the inverse of the invariant minimal volume of space, cf. (\ref{omegaop}). If we assume that the scale of $\varkappa$ is Planckian, then the emerging cosmological constant will also be Planckian. It reproduces the usual 122 orders of magnitude error resulting from the identification of the cosmological constant with the zero-point energy integrated over all the momenta of virtual vacuum excitations below the (believed to be) natural cutoff at the Planck scale. However, this assumption is not confirmed by our estimation of the mass gap in quantum SU(2) gauge theory with the coupling constant $g$ \cite{my-ymmg}, where the scale of $\varkappa$ is found to be close to the scale of the mass gap $\Delta \mu \sim (g^2\hbar^4 \varkappa)^{1/3}$. The latter is expected to be in the range of the meson masses in QCD or the glueball masses in pure SU(3) gauge theory, i.e. around 1 GeV plus-minus one order of magnitude. By replacing the Planckian 
$\varkappa \sim (\hbar G)^{-3/2} $ with a GeV-scale $\varkappa$ as suggested by the mass gap estimation in \cite{my-ymmg} we obtain the value of $\lambda$ which is $3^2 \times (19 \pm 1)$ orders of magnitude smaller than Planckian, i.e. roughly consistent with the observable value of the cosmological constant. Note that this observation is quantitatively consistent with the similar estimation of the cosmological constant in our previous work on the precanonical quantization of vielbein general relativity \cite{ikv1,ikv2,ikv3,ikv4,ikv5}.

\subsection*{Acknowledgment} I gratefully acknowledge V.A. Kholodnyi for his interest, many inspirational discussions,  and  insightful comments. 
I also thank 
M. Wright for his careful reading of the manuscript and suggested improvements. 


\end{document}